\documentclass[final,5p,times,twocolumn]{elsarticle}
\usepackage{subfigure}
\usepackage{atlasphysics}
\usepackage{preprintcover}
\newcommand{\ptdijet}{\ensuremath{p_{\mathrm T}^{\mathrm{dijet}}}}

\biboptions{sort&compress}
\def\pythia{{\sc Pythia 8}}
\def\sherpa{{\sc Sherpa}}

\def\herwig{{\sc Herwig}}
\def\herwigplus{{\sc Herwig++}}

\def\PZ{Z}
\def\PW{W}
\def\Pbeauty{b}
\def \APbeauty{\overline{b}}
\def\Pcharm{c}
\def \APcharm{\overline{c}}
\def\Ptop{t}
\def \APtop{\overline{t}}
\def\zbb{\mbox{$\PZ\hspace*{-0.5 mm}\to\hspace*{-0.5 mm}\Pbeauty\APbeauty$}}
\def\zcc{\mbox{$\PZ\hspace*{-0.5 mm}\to\hspace*{-0.5 mm}\Pcharm\APcharm$}}
\def\wqq{\mbox{$\PW\hspace*{-0.5 mm}\to\hspace*{-0.5 mm}q\overline{q}^\prime$}}

\def\ttbar{$\Ptop\APtop$}
\def\split{$\cal{S}_{NN}$}

\def\NZ{$N_{\PZ\rightarrow\Pbeauty\APbeauty}$}

\def\hbb{\mbox{$H\hspace*{-0.5 mm}\to\hspace*{-0.5 mm}\Pbeauty\APbeauty$}}

\def\zbbxsec{$\sigma^{\mathrm{fid}}_{\PZ\rightarrow\Pbeauty\APbeauty}$}
\def\acc{${\cal{C}}_{\PZ\rightarrow\Pbeauty\APbeauty}$}

\PreprintCoverPaperTitle{Measurement of the cross section of high transverse momentum \zbb~production \\ in proton--proton collisions at $\sqrt{s}=8$\,\TeV\ with the ATLAS Detector}  
\PreprintIdNumber{CERN-PH-EP-2014-064}  
\PreprintCoverAbstract{This Letter reports the observation of a high transverse momentum \zbb~signal in proton--proton collisions at $\sqrt{s}=8$\,\TeV\ and the measurement of 
its production cross section. The data analysed were collected in 2012 with the ATLAS detector at the LHC and correspond to an integrated 
luminosity of 19.5\,fb$^{-1}$. The \zbb~decay is reconstructed from a pair of $b$-tagged jets, 
clustered with the anti-$k_t$ jet algorithm with $R = 0.4$,
that have low angular separation and form a dijet 
with $\pt > 200$\,\GeV. The signal yield is extracted  from a fit to the
dijet invariant mass distribution, with the dominant, multi-jet background mass shape estimated by employing a fully data-driven technique that reduces the
dependence of the analysis on simulation. The fiducial cross section is determined to be
$$
  \sigma^\mathrm{fid}_{\PZ\rightarrow\Pbeauty\APbeauty} = 2.02\pm0.20~({\rm stat.})
  ~\pm0.25~({\rm syst.)}\pm 0.06~({\rm lumi.})~\,{\rm pb}  =
  2.02 ~\pm{0.33}~\,{\rm pb}\,,
$$
in good agreement with next-to-leading-order theoretical predictions.  }  
\PreprintJournalName{Physics Letters B}  

\renewcommand*{\today}{8th September, 2014}

\begin{document}

\begin{frontmatter}

\title{ Measurement of the cross section of high transverse momentum \zbb~production \\ in proton--proton collisions at $\sqrt{s}=8$\,\TeV\ with the ATLAS Detector}

\author{ATLAS Collaboration}

\begin{abstract}
This Letter reports the observation of a high transverse momentum \zbb~signal in proton--proton collisions at $\sqrt{s}=8$\,\TeV\ and the measurement of 
its production cross section. The data analysed were collected in 2012 with the ATLAS detector at the LHC and correspond to an integrated 
luminosity of 19.5\,fb$^{-1}$. The \zbb~decay is reconstructed from a pair of $b$-tagged jets, 
clustered with the anti-$k_t$ jet algorithm with $R = 0.4$,
that have low angular separation and form a dijet 
with $\pt > 200$\,\GeV. The signal yield is extracted  from a fit to the
dijet invariant mass distribution, with the dominant, multi-jet background mass shape estimated by employing a fully data-driven technique that reduces the
dependence of the analysis on simulation. The fiducial cross section is determined to be
$$
  \sigma^\mathrm{fid}_{\PZ\rightarrow\Pbeauty\APbeauty} = 2.02\pm0.20~({\rm stat.})
  ~\pm0.25~({\rm syst.)}\pm 0.06~({\rm lumi.})~\,{\rm pb}  =
  2.02 ~\pm{0.33}~\,{\rm pb}\,,
$$
in good agreement with next-to-leading-order theoretical predictions. 
\end{abstract}

\begin{keyword}
LHC \sep boosted $\Pbeauty\APbeauty$ topologies 
\end{keyword}

\end{frontmatter}

\section{Introduction}

High transverse momentum (\pt), hadronically decaying, electroweak-scale bosons have already been used in searches at the 
LHC~\cite{Aad:2013wxa, CMS:DijetWZ, CMS:SemilepWZ, Aad2012369, Chatrchyan2012284}, and are expected to play an increasingly significant role as the LHC 
moves to higher centre-of-mass energies in 2015. 
Therefore it is important to study them directly.
This Letter presents the observation of a high-\pt~\zbb~signal in a fully hadronic final state, and a
measurement of its production cross section.
The measurement is compared to the next-to-leading-order (NLO) matrix element plus parton-shower predictions 
of POWHEG~\cite{Alioli:2010qp, Nason:2004rx, Frixione:2007vw, Alioli:2010xd} and aMC@NLO~\cite{amcatnlo}, 
where the parton-shower, hadronisation and underlying-event modelling are provided by 
{\sc{Pythia}}-8.165~\cite{ref:pythia8} and \herwigplus~\cite{Bahr:2008pv} respectively. 
This first measurement of a high-\pt\ electroweak-scale boson in an all-hadronic final state at the LHC demonstrates the validity of both the analysis techniques used and of the state-of-the-art NLO plus parton-shower particle-level predictions for electroweak-scale bosons decaying to $\Pbeauty\APbeauty$. It is therefore of great relevance for the search 
for the \hbb~signal in the (most sensitive) high Higgs boson \pt~range~\cite{Aad:2012gxa}, 
as well as for searches for \TeV-scale resonances decaying to $\Pbeauty\APbeauty\Pbeauty\APbeauty$ via $ZZ$, $ZH$ 
or $HH$~\cite{PhysRevD.88.114005, Gouzevitch:2013qca}. 
A high-\pt~\zbb~signal can also provide a useful benchmark for validating the performance of the ATLAS detector (for example, the $b$-jet energy scale\footnote{The use of the $\PZ\hspace{-0.5 mm}\to\hspace{-0.5 mm}\Pbeauty\APbeauty$ peak to constrain 
the $b$-jet energy scale at the Tevatron experiments was demonstrated in Ref. \cite{Donini:2008nt}.}); and for testing and optimising analysis methods relevant for physics studies involving high-\pt~jets that contain $b$-hadrons ($b$-jets).

The analysis described in this Letter is designed to select $\Pbeauty\APbeauty$ decays of $Z$-bosons with $\pt > 200$\,\GeV,
in proton--proton collisions at $\sqrt{s} = 8$\,\TeV. The high-\pt~requirement helps to enhance the signal relative to $\Pbeauty\APbeauty$
production in multi-jet events (predominantly gluon splitting to $\Pbeauty\APbeauty$ in this high-\pt~regime), 
which is the dominant source of background and has a
more steeply-falling \pt~spectrum.  In order to minimise
the dependence on simulation, the analysis employs a fully
data-driven technique for the determination of the invariant mass
spectrum of the multi-jet background. This is especially
important given that Monte Carlo (MC) generators have not been
tested thoroughly in this region of the $\Pbeauty\APbeauty$ production phase space.

\section{The ATLAS detector}
ATLAS is a multi-purpose particle physics experiment~\cite{ref:AtlasDet} at the LHC. The detector 
layout\footnote{ATLAS uses a right-handed coordinate system with its origin at the nominal interaction point (IP) in the centre of the 
detector and the $z$-axis along the beam pipe. The $x$-axis points from the IP to the centre of the LHC ring, and the y-axis points 
upwards. Cylindrical coordinates $(r, \phi)$ are used in the transverse plane, $\phi$ is the azimuthal
angle around the beam pipe. The pseudorapidity, $\eta$, is defined in terms of the polar angle $\theta$ as $\eta = -\ln[\tan (\theta/2)]$.}
consists of inner tracking devices surrounded by a superconducting solenoid, electromagnetic
and hadronic calorimeters, and a muon spectrometer. 
The inner tracking system provides charged-particle tracking in the pseudorapidity region $|\eta| < 2.5$ and vertex reconstruction. It consists of a silicon pixel detector, a silicon microstrip
tracker, and a straw-tube transition radiation tracker. The system is surrounded by a solenoid that produces a 2\,T axial magnetic
field.
The central calorimeter system consists of a liquid-argon electromagnetic (EM) sampling calorimeter
with high granularity and an iron/scintillator-tile calorimeter providing hadronic energy measurements in
the central pseudorapidity range ($|\eta| < 1.7$). The endcap and forward regions are instrumented with liquid-argon
calorimeters for both electromagnetic and hadronic energy measurements up to $|\eta| = 4.9$. The muon spectrometer
is operated in a magnetic field provided by air-core superconducting toroids and includes tracking
chambers for precise muon momentum measurements up to $|\eta| = 2.7$ and trigger chambers covering the range $|\eta| < 2.4$.
A three-level trigger system is used to select interesting events~\cite{Aad:2012xs}.
The Level-1 trigger reduces the event rate to below 75\,kHz using hardware-based trigger
algorithms acting on a subset of detector information. Two software-based trigger levels, referred to collectively as the High-Level Trigger (HLT),
further reduce the event rate to about 400\,Hz using information from the entire detector.

\section{Data, simulated samples, and event reconstruction}
The data sample used in this analysis, after requiring that certain quality criteria be satisfied, corresponds to an integrated luminosity 
of ${\cal L}=19.5\pm0.5$\,fb$^{-1}$, and was recorded by ATLAS in 2012. The uncertainty on the integrated luminosity is derived, 
following the same methodology as that detailed in Ref. \cite{Aad:2013ucp}, from a calibration of the luminosity scale using 
beam-separation scans performed in November 2012.

MC event samples simulated with the GEANT4-based~\cite{ref:Geant4}
ATLAS detector simulation~\cite{Collaboration:2010wq} are used to
model the \zbb~signal and the small \ttbar, \zcc~and \wqq~background
contributions. In addition, multi-jet MC event samples are used for
studying the trigger modelling in simulation. The effect of multiple proton-proton interactions in the same bunch crossing (pile-up) is included in the simulation.

The \zbb~signal is modelled using \sherpa-1.4.3~\cite{Gleisberg:2008ta}, with the CT10~\cite{Lai:2010vv} NLO 
parton distribution function (PDF) set. An alternative \zbb\ model was generated with {\sc{Pythia}}-8.165~\cite{ref:pythia8} and the CTEQ6L1~\cite{Pumplin:2002vw} 
leading-order (LO) PDF set and is used to determine the systematic uncertainty associated with \zbb\ modelling. The \zcc~background is also generated 
with \sherpa-1.4.3 and the CT10 PDF set. The \ttbar~background 
is simulated with MC@NLO-4.06~\cite{ref:MCatNLO1} interfaced to \herwig-6.520~\cite{ref:Herwig} for the fragmentation 
and hadronisation processes, including {\sc{Jimmy}}-4.31~\cite{Butterworth:1996zw} for the underlying-event description. The
top quark mass is fixed at 172.5\,\GeV, and the PDF set CT10 is
used. The \wqq~and multi-jet MC samples are generated using {\sc{Pythia}}-8.165
with the CT10 PDF set.

Jets are reconstructed using the anti-$k_t$ jet clustering
algorithm~\cite{Cacciari:2008gp}, with radius parameter $R = 0.4$\,. The
inputs to the reconstruction algorithm are topological calorimeter
cell clusters~\cite{Aad:2011he} calibrated at the EM energy scale. The effects of pile-up on 
jet energies  are accounted for by a jet-area-based correction~\cite{Cacciari:2008gn}.
Jets are then calibrated to the hadronic energy scale using \pt- and
$\eta$-dependent calibration factors based on MC simulations
and the combination of several {\it in situ} techniques applied to
data~\cite{Aad:2011he}. To remove jets with a significant contribution
from pile-up interactions, it is required that
at least 50\% of the summed scalar \pt~of tracks
matched to a jet belongs to tracks originating from the primary 
vertex.\footnote{Amongst all reconstructed proton-proton collisions in a bunch crossing, the primary vertex is defined as the vertex with the highest 
summed track $\pt^2$.}

The flavour of jets in the ATLAS simulation is defined by matching jets to hadrons
with $\pt>5$\,\GeV. A jet is labelled as a $b$-jet if a $b$-hadron is found within
$\Delta R = \sqrt{(\Delta \eta)^2 + (\Delta \phi)^2}=0.3$ of the jet axis; otherwise, if a
$c$-hadron is found within the same distance the jet is labeled as a $c$-jet; and if neither is the case then the jet is labelled as
a light (quark) jet.
The lifetime and other properties of $b$-hadrons are used to identify ($b$-tag)
$b$-jets with $|\eta| < 2.5$\,, 
by exploiting the properties and topology of their decay products,  such as the impact parameter of tracks (defined as a track's distance of
closest approach to the primary vertex in the transverse plane), the presence of displaced vertices, and the reconstruction of 
$c$-hadron and $b$-hadron decays. The $b$-tagging algorithm used in this analysis~\cite{ATLAS-CONF-2012-097} combines the above information
using multivariate techniques and is configured to achieve an efficiency of 70\% for tagging 
$b$-jets in a MC sample of \ttbar~events, while rejecting 80\% of $c$-jets and more than 99\% of light jets in the same sample.

\section{Event selection}
The events of interest in this analysis were
triggered by a combination of six jet-based triggers. The 
most efficient of these triggers (accepting about 70\% of the signal events) requires two jets identified as $b$-jets by a dedicated HLT 
$b$-tagging algorithm, 
with transverse energies ($E_{\mathrm T}$) above 35\,\GeV, and a jet with $E_{\mathrm T}> 145$\,\GeV\ that may or 
may not be one of the $b$-tagged jets. The trigger efficiency for the \sherpa~signal events passing the full offline event selection is $88.1\%$.

The event selection requires that there be at least three but no more than five jets
with $|\eta| <2.5$ and $\pt > 30$\,\GeV, and that exactly two of them be $b$-tagged. 
The $b$-tagged jets must each have $\pt > 40$\,\GeV. 
The angular distance, $\Delta R$, between
them must be smaller than 1.2 and the transverse momentum of the dijet system they form, $\ptdijet$, must be greater than $200$\,\GeV.

\begin{figure}[!]
\begin{center}
\subfigure[]{\includegraphics[width=0.48\textwidth]{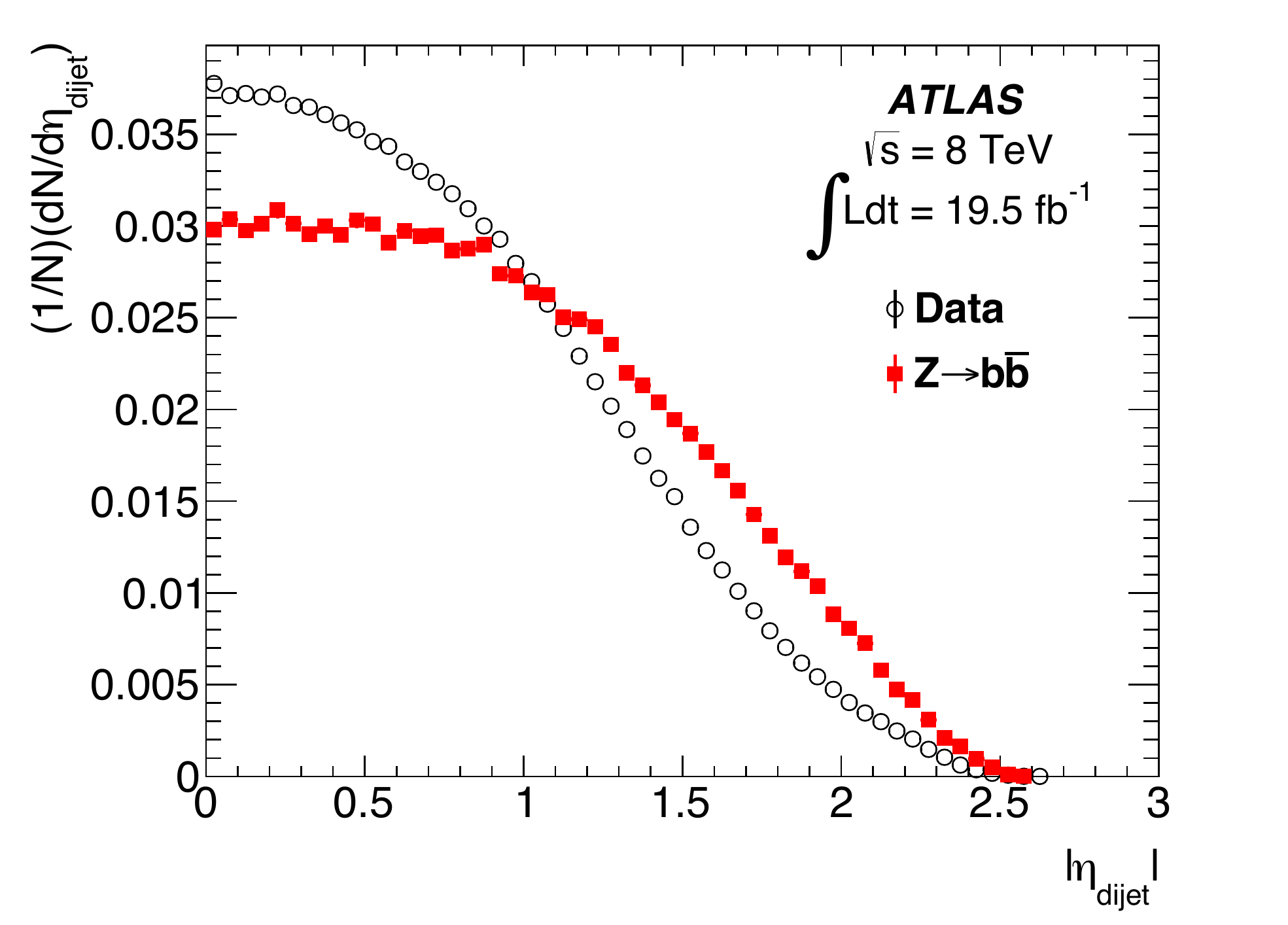}}
\subfigure[]{\includegraphics[width=0.48\textwidth]{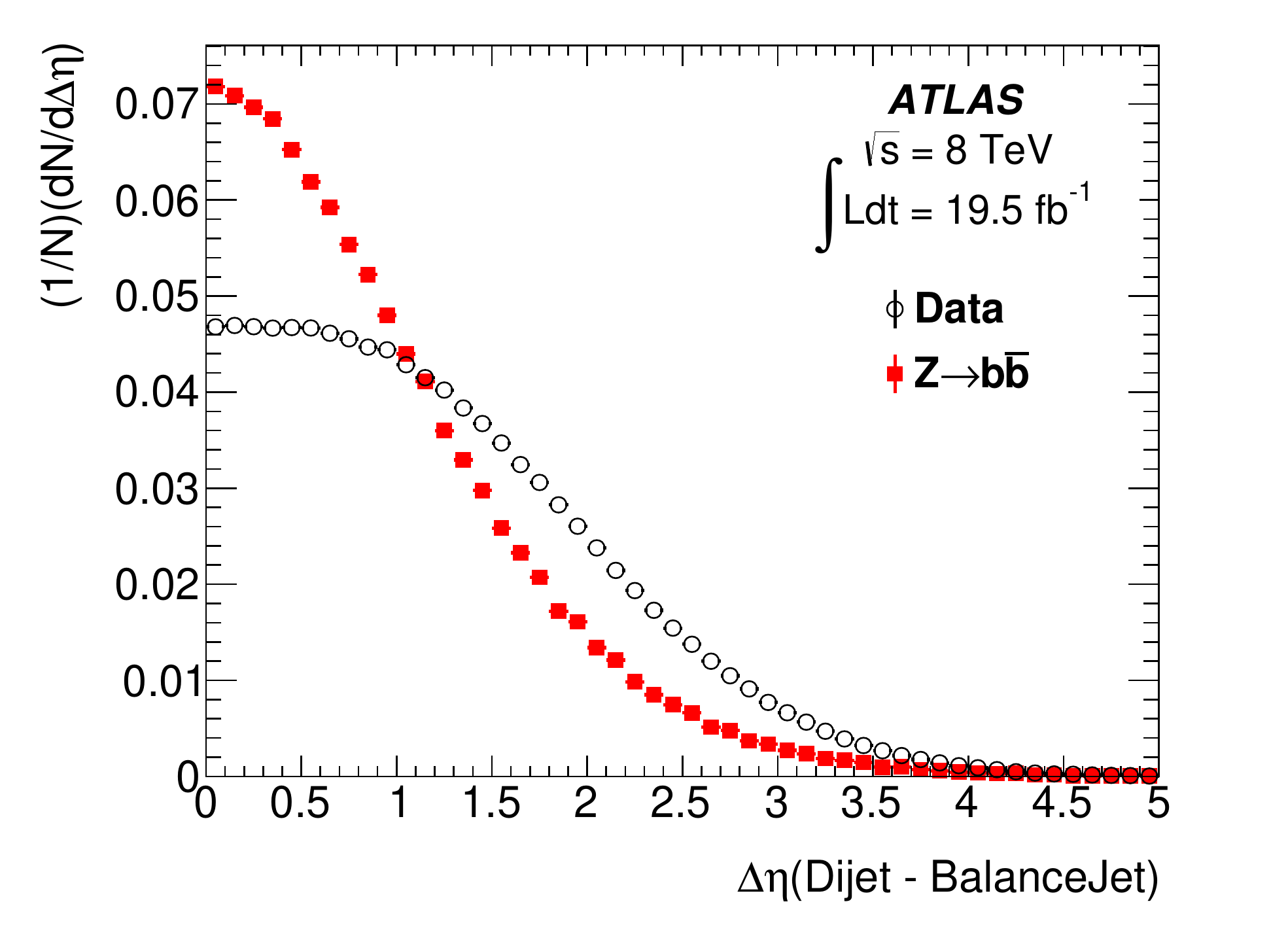}}
\subfigure[]{\includegraphics[width=0.48\textwidth]{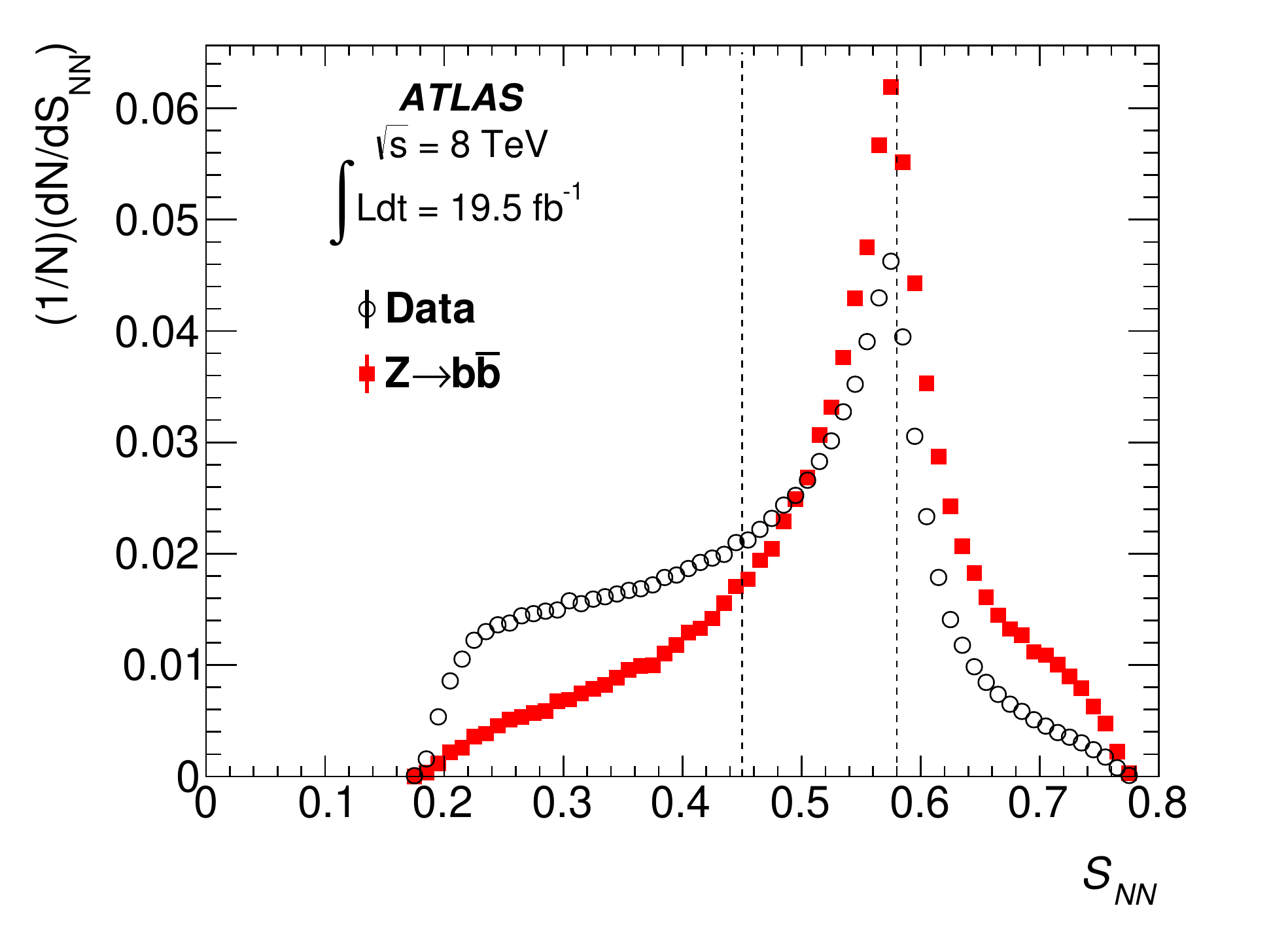}}
\caption{The distributions of: (a) the dijet pseudorapidity, $|\eta_{\mathrm{dijet}}|$;
(b) the pseudorapidity difference, $|\Delta \eta|$, between the dijet and the balancing jet; and (c) the neural network discriminant \split,
 in the \zbb~signal (red squares) and in the data (black circles), including all events with  $60<m_{\mathrm{dijet}}<160$\,\GeV. The data is dominated by the multi-jet background.
 The two dashed lines in (c) indicate the \split~values defining the Signal (\split\,$>0.58$) and Control (\split\,$<0.45$) Regions.}
\label{fig:splitvars}
\end{center}
\end{figure}

The final step of the event selection uses two variables with
significant discrimination between signal and background to define two sets of events, one signal-enriched and the other signal-depleted, 
referred to hereafter as ``Signal Region'' and ``Control Region''. The two variables, which are
combined with an artificial neural network (ANN) into a single discriminant,
\split, are: (1) the dijet pseudorapidity, $\eta_{\mathrm{dijet}}$;
and (2) the pseudorapidity difference, $\Delta \eta$, between the dijet and the balancing jet, where the 
balancing jet is chosen to be the one that, when added to the dijet, gives the three-jet system with the smallest transverse momentum. The correlation of these two variables with the dijet invariant mass, $m_{\mathrm{dijet}}$, is very small, allowing the ANN to be trained using selected data events outside the mass window [80, 110]\,\GeV\ as background and \zbb\ MC events as signal. 
Fig.~\ref{fig:splitvars} depicts the distributions of $\eta_{\mathrm{dijet}}$, $\Delta \eta$ and \split\ in the signal MC sample and in the data.
The data shown here include all events with $60<m_{\mathrm{dijet}}<160$\,\GeV, and are representative of the background as the signal contribution is 
estimated to be only about 1\%.
The Signal Region is defined by \split\,$>0.58$ and the Control Region by \split\,$<0.45$. 
The discriminating power of $\eta_{\mathrm{dijet}}$ and $\Delta \eta$ stems from the fact that signal production proceeds predominantly 
via a quark--gluon hard scatter, as opposed to the dominant multi-jet background which is largely initiated by gluon-gluon scattering. 
Due to the differences between the gluon and quark PDFs, the $Z$\,$+$\,jet system tends to be more boosted along the beam
axis; hence the $Z$-boson is produced with higher $\eta$ and smaller $\Delta\eta$ separation from its recoil, compared to the background.

\begin{figure}[ht]
\begin{center}
\includegraphics[width=0.49\textwidth]{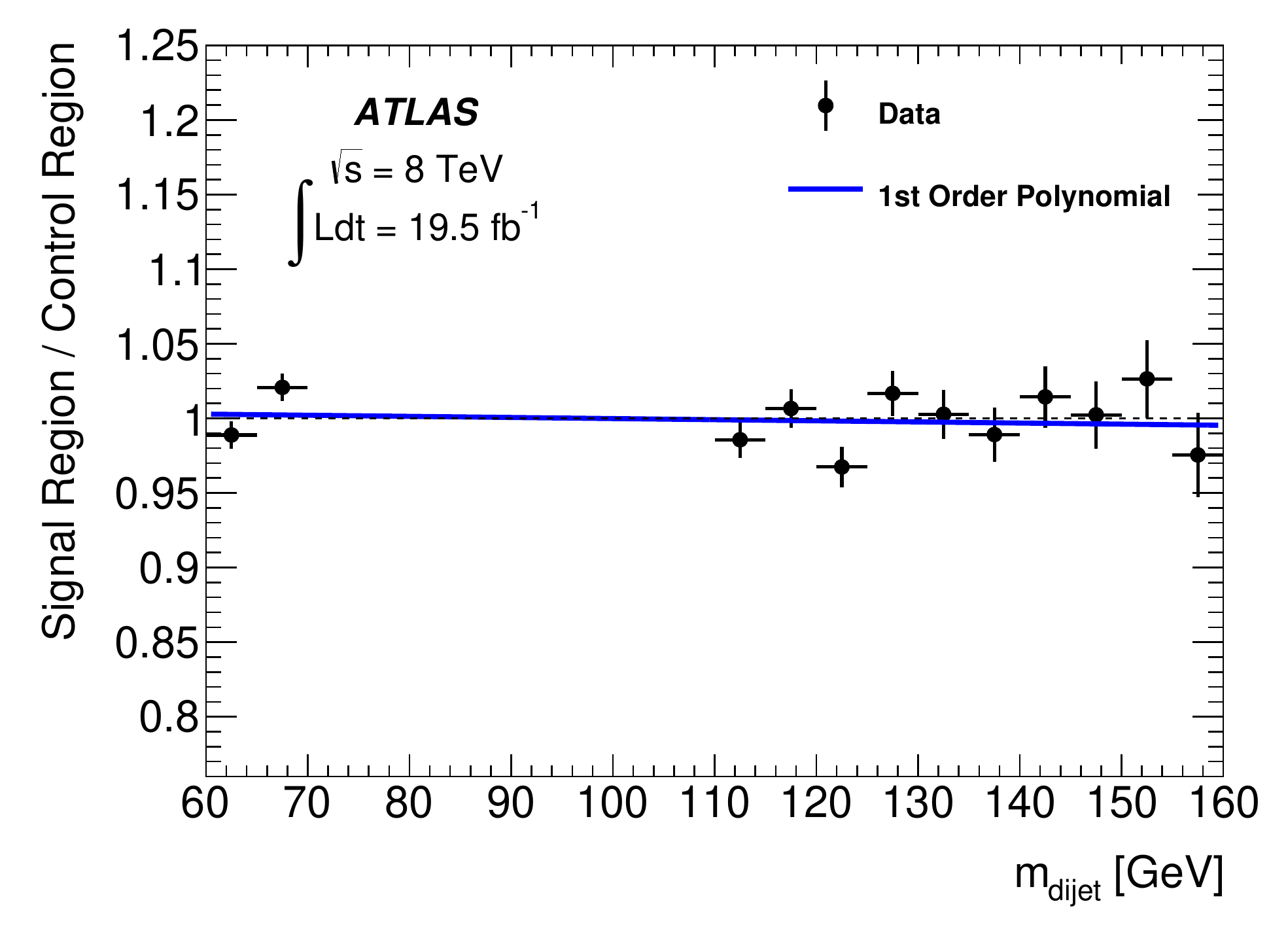}
\caption{The normalised ratio of dijet mass distributions in the Signal and Control Regions, excluding the signal mass window, fitted with a first-order polynomial. 
The dashed line indicates unity.
}
\label{fig:signal-vs-control}
\end{center}
\end{figure}

Since \split~is minimally correlated with $m_{\mathrm{dijet}}$
the Control Region can be used as an unbiased
model of the background in the Signal Region. 
Fig.~\ref{fig:signal-vs-control} shows the normalised ratio of the $m_{\mathrm{dijet}}$ distributions in the Signal
and Control Regions, excluding the $Z$ mass window. A first-order polynomial fit to this distribution gives a good $\chi^2$ 
probability, 0.18, and a gradient consistent with zero,  $(-1.37\pm1.10)\times 10^{-4}$\,\GeV$^{-1}$. 
In addition, the validity of assuming minimal correlation is supported by a test, performed with events from a \pythia~multi-jet MC sample satisfying the above analysis requirements, giving a ratio of the above distributions consistent with being flat. 
The impact of possible differences in the background $m_{\mathrm{dijet}}$ shape between the Signal and Control Regions 
is considered as one of the systematic uncertainties on the measurement.

The total number of data events satisfying the full analysis selection is
236172 in the Signal Region and 474810 in the Control Region. The
signal-to-background ratio in a 30\,\GeV\ window
around the $Z$-boson mass is expected to be about 6\% (2\%) in the Signal (Control) Region.
The \ttbar~events are estimated to represent about 0.5\% of the total background in both the Signal and Control Regions, 
and the \zcc~and \wqq~backgrounds are approximately 8\% and 6\% of the signal, respectively.

\section{Cross-section definition}
\label{sec:xsec}
The fiducial cross section of resonant $Z$-boson production, with $Z$ decaying to $\Pbeauty\APbeauty$,
$\sigma^{\mathrm{fid}}_{Z\rightarrow\Pbeauty\APbeauty}$, is defined as
follows. Particle-level jets in MC \zbb~events are reconstructed from stable particles (particles with
lifetime in excess of 10\,ps, excluding muons and neutrinos) using the anti-$k_t$ algorithm
with radius parameter $R = 0.4$\,. 
There must be two particle-level $b$-jets in the event that satisfy the following fiducial
conditions: $\pt > 40$\,\GeV, $|\eta| < 2.5$ for the individual jets; and $\Delta R({\mathrm{jet1,jet2}})<1.2$,
$\pt^\mathrm{dijet}>200$\,\GeV, $60<m_{\mathrm{dijet}}<160$\,\GeV\ for the dijet system. 

The cross section is extracted from the measured yield of \zbb~events in the data, \NZ, as\vspace*{-2mm}

$$
\sigma^{\mathrm{fid}}_{Z\rightarrow\Pbeauty\APbeauty} = 
\frac{N_{\PZ\to\Pbeauty\APbeauty}}{{\cal{L}} \cdot {\cal{C}}_{\PZ\to\Pbeauty\APbeauty}}\,,
$$
where ${\cal{C}}_{Z\rightarrow\Pbeauty\APbeauty}$ is the efficiency correction factor to correct the detector-level 
\zbb~yield to the particle level. The value of ${\cal{C}}_{Z\rightarrow\Pbeauty\APbeauty}$
in the \sherpa~MC signal is found to be 16.2\%, which can be factorised into the product of: trigger efficiency (88.1\%), $b$-tagging and
 kinematic selection efficiency (52.7\%), and the efficiency of the \split~requirement that defines the Signal Region (35.0\%). The uncertainties on ${\cal{C}}_{Z\rightarrow\Pbeauty\APbeauty}$ are discussed in Section~\ref{sec:syst}.

\section{Signal extraction}
The signal yield is extracted by fitting simultaneously the $m_{\mathrm{dijet}}$ distributions of the Signal and Control Regions 
in the range $[60,160]$\,\GeV\ with a binned, extended maximum-likelihood (EML) fit, using a bin width of 1\,\GeV.

The $Z\rightarrow\Pbeauty\APbeauty$ signal shape is
modelled in the EML fit as the sum of three Gaussians. This empirical model describes well both the \sherpa\ and the {\sc{Pythia}}~signal MC samples, albeit with slightly different parameters. 
Given this, the \sherpa-based model is used as the baseline for the fit, and the {\sc{Pythia}}-based model is used for an estimate of the systematic uncertainty on the measurement due to the signal shape.
The only free parameters of the signal model in the EML fit are 
the yield in the Signal Region and the shift, $\delta M_Z$, of the mean of the
narrowest Gaussian from its MC-predicted value. The widths and relative contributions of the three Gaussians, as well as the differences 
between the mean of each of the two wider Gaussians and the narrowest one, are fixed to
the values determined by a separate fit to the signal MC $m_{\mathrm{dijet}}$ distributions. 
Given that the Control Region is not signal-free, the simultaneous fit includes
a signal component in both the Signal Region and the Control Region. The relative proportion of signal in the two regions,
$
R_Z = (N_{\PZ\rightarrow\Pbeauty\APbeauty}^{\rm{Control}})/(N_{\PZ\rightarrow\Pbeauty\APbeauty}^{\rm{Signal}})
$,
is fixed to the value predicted by \sherpa, $R_Z=0.62$. This choice is supported by the good agreement found between \sherpa~and data
in the \split~distribution obtained from a pure sample of high-\pt~$Z\rightarrow\mu^+\mu^-$ events, as discussed in Section~\ref{sec:syst}.

The dominant multi-jet background is modelled in the EML fit using a seventh-order Bernstein polynomial \cite{Bernstein1912}.
This is purely an empirical model and the order of the polynomial is chosen by a $\chi^2$ probability saturation test 
by fitting the $m_{\mathrm{dijet}}$ distribution in the Control Region
with the background-only hypothesis and an increasing polynomial order, until no improvement is seen 
in the fit quality when adding higher-order terms.
The coefficients of the Bernstein polynomial are determined by the simultaneous EML
fit and are identical for the Signal and Control Regions. In this way,
the signal-depleted Control Region constrains the background
prediction in the Signal Region. 
The only additional parameters of the fit are the background normalisations in the 
Signal and Control Regions. 
\begin{figure*}[ht]
\begin{center}
\subfigure[]{\includegraphics[width=0.49\textwidth]{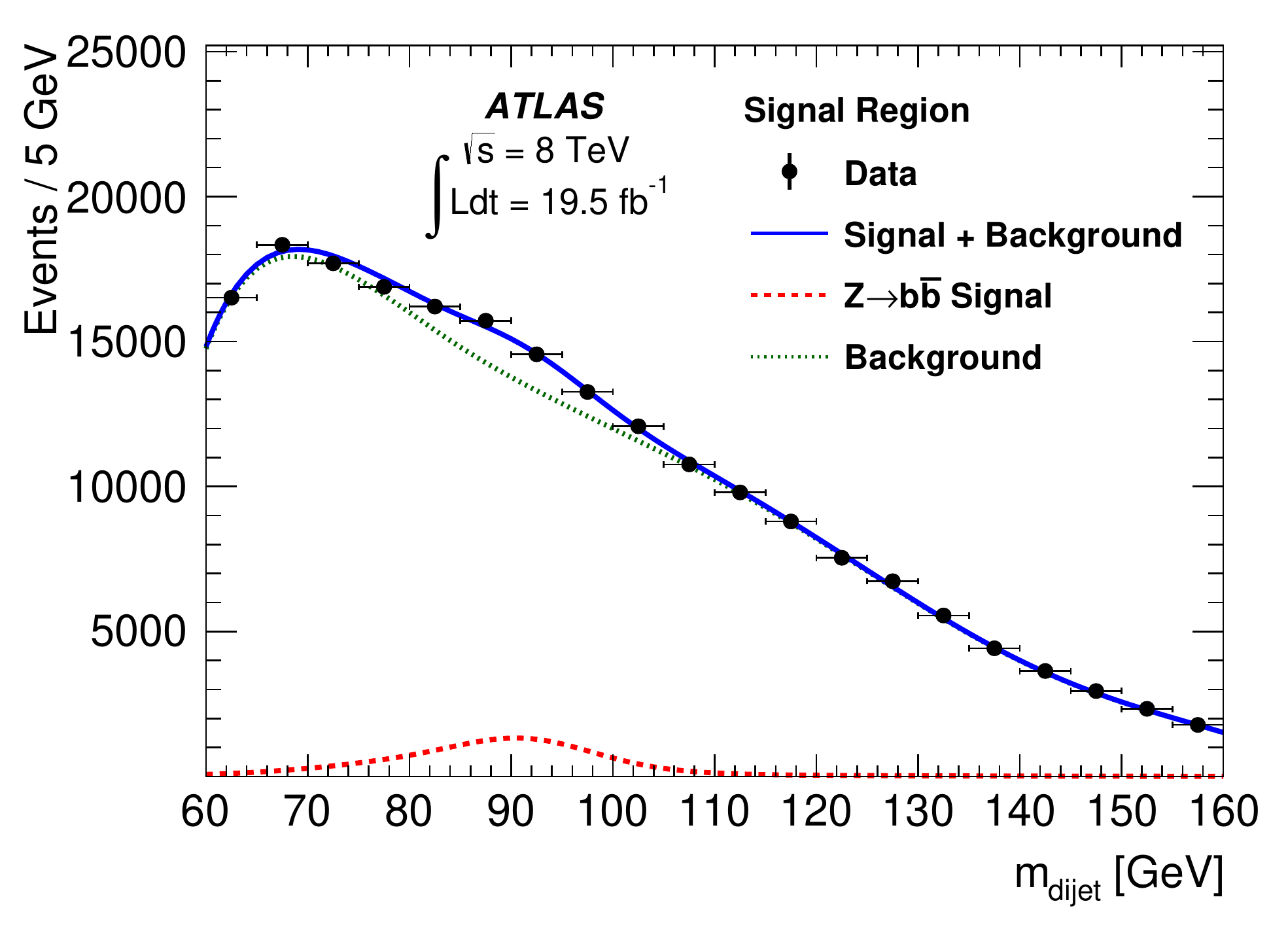}}
\subfigure[]{\includegraphics[width=0.49\textwidth]{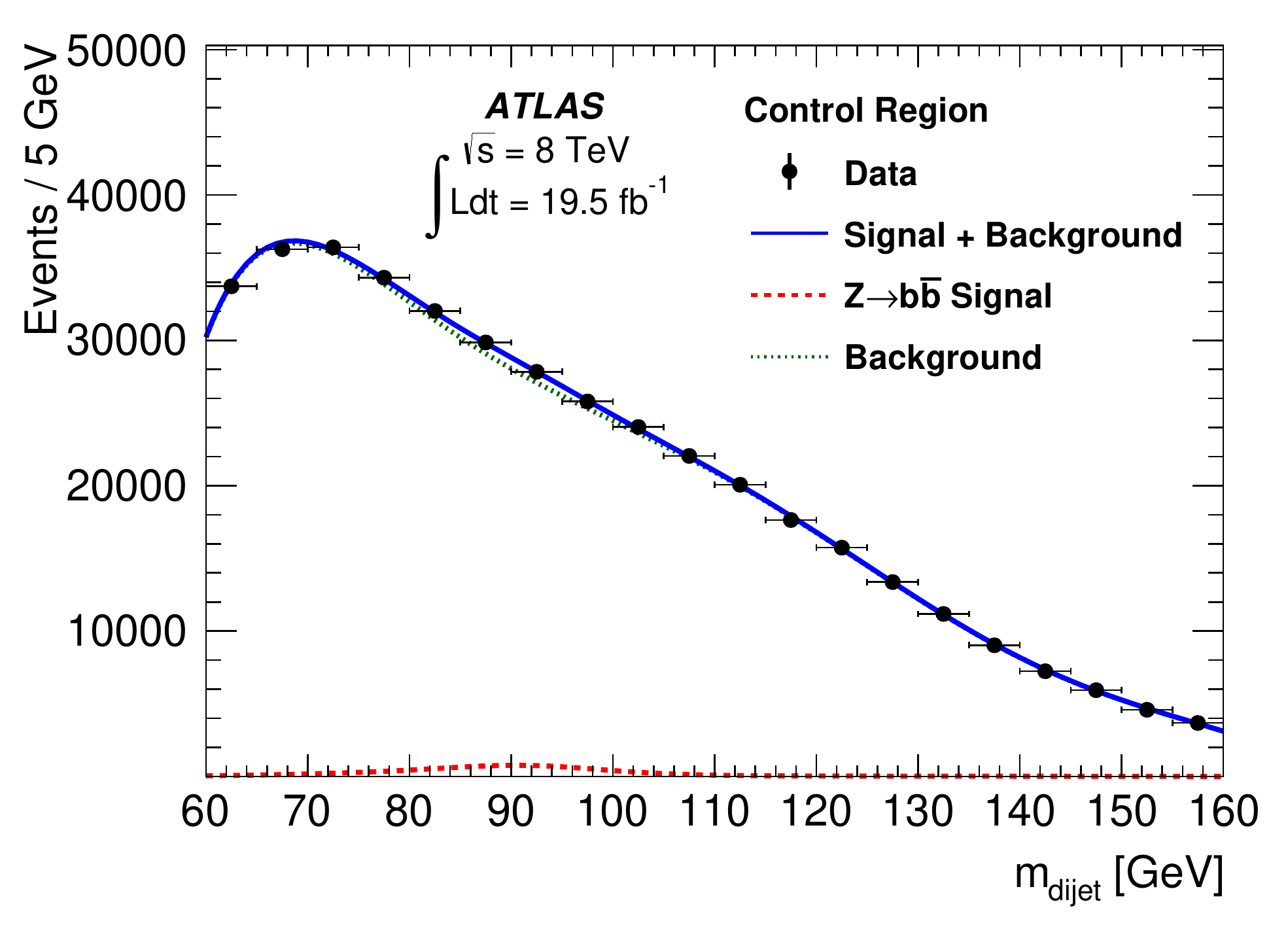}}
\subfigure[]{\includegraphics[width=0.49\textwidth]{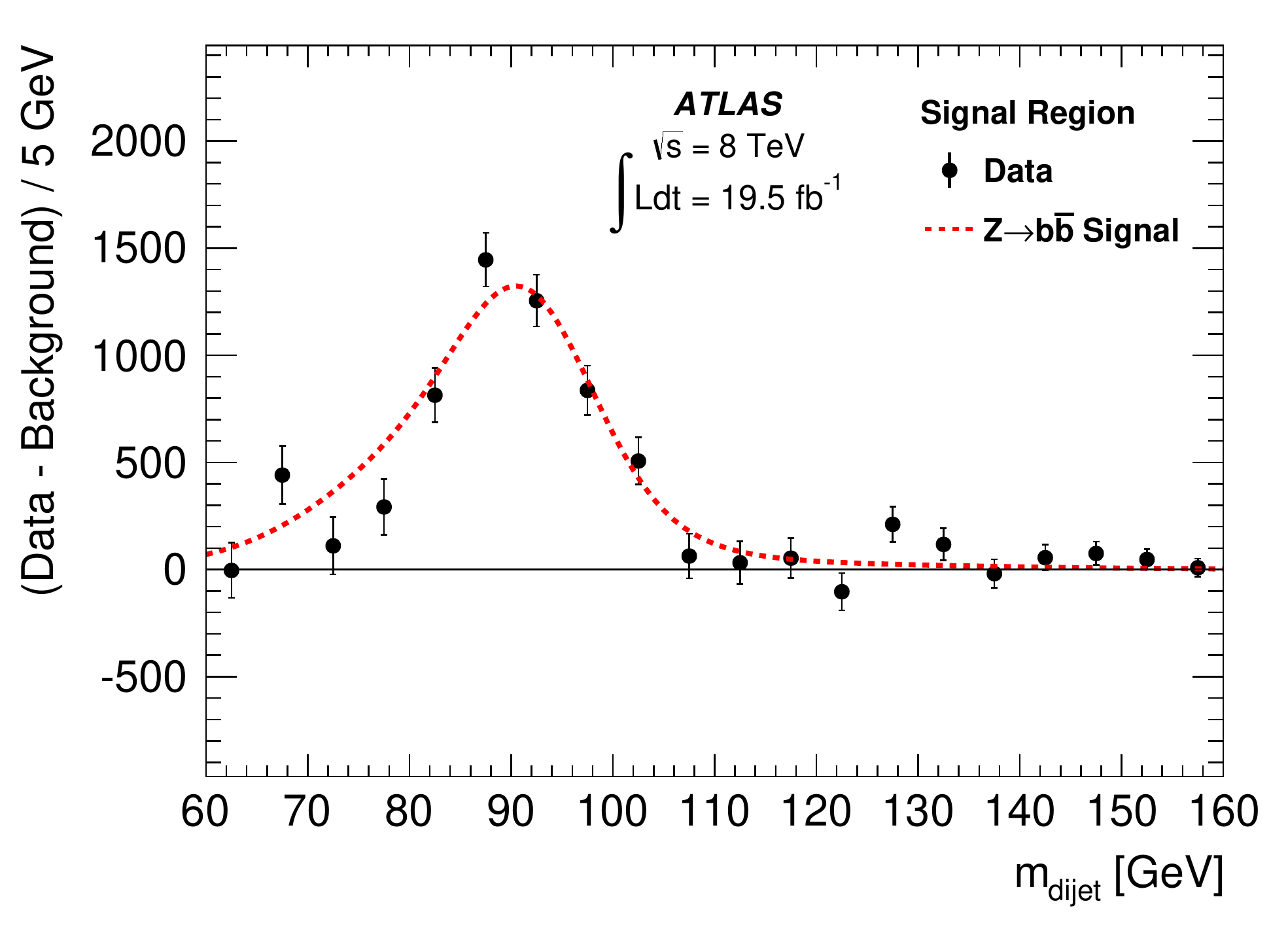}}
\subfigure[]{\includegraphics[width=0.49\textwidth]{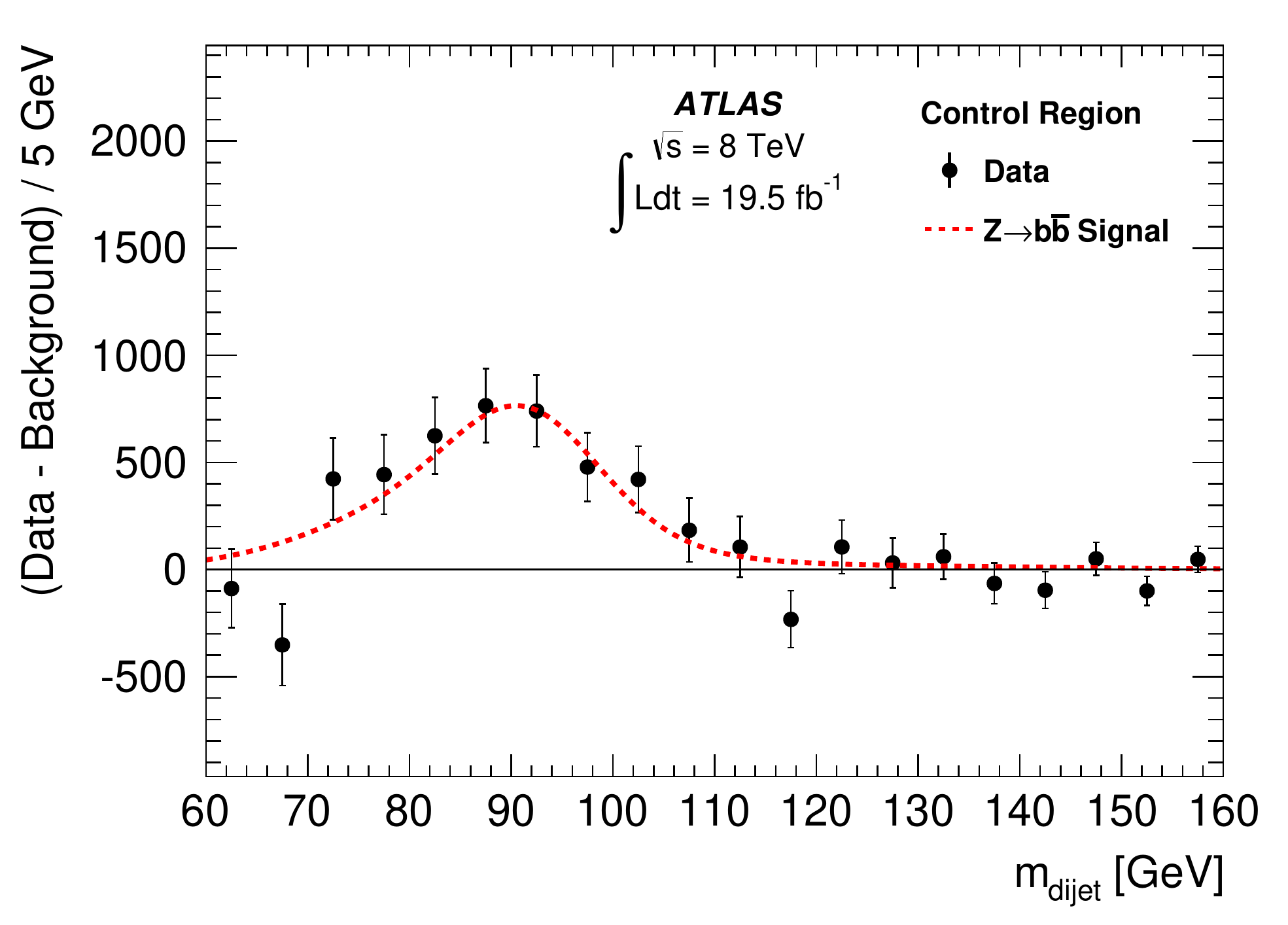}}
\caption{The result of the simultaneous extended maximum likelihood fit to the dijet mass distributions in (a) the Signal Region and (b) the Control Region, and the corresponding background-subtracted distributions
(c) and (d), using the \sherpa~signal model. The lines represent the signal (dashed), backgrounds (dotted) and the sum of the two (solid).
}
\label{fig:fit-results}
\end{center}
\end{figure*}

The small \zcc, \ttbar~and \wqq~backgrounds are included as separate components in the EML fit for both the Signal and Control Regions, with their 
$m_{\mathrm{dijet}}$ shapes being parameterised from MC simulation as follows. The \zcc~and \wqq~components are each modelled as three-Gaussian sums like the 
signal, with all parameters fixed to values from fits to MC simulation. The means of the Gaussians are expressed with respect to the mean of the 
narrowest \zbb~Gaussian: this couples the position of these backgrounds to the \zbb~signal. The \wqq~component is normalised absolutely to its {\sc{Pythia}} 
LO cross section, corrected to NLO by a $K$-factor derived using MCFM~\cite{Campbell:2010ff}. The acceptance of the \zcc~background is taken from the simulation, but its yield is linked to the fitted \zbb~yield, since the \zcc~production differs from the signal only in the well-known branching fractions of the $Z$ decays. 
All properties of the \ttbar~component are fixed using the \ttbar~simulation, with normalisation from the NNLO+NNLL
prediction of the \ttbar~production cross section~\cite{Cacciari:2011hy,Baernreuther:2012ws,Czakon:2012zr,Czakon:2012pz,Czakon:2013goa,Czakon:2011xx}.
The contribution from Higgs decays to $\Pbeauty\APbeauty$ is expected to be $\sim3\%$ of the \zbb\ signal and localised away from the signal peak: therefore no such component is included in the EML fit.

The fit procedure has been validated using a comprehensive set of tests based on pseudo-experiments, which have demonstrated that the yield and its uncertainty are accurately determined by the fit procedure over a wide range of input signal yields. In particular, the fit procedure is robust against fitting artefacts like false dips or peaks: a consequence of fitting both signal and control regions simultaneously, with the ratio of \zbb\ in each region fixed.

Fig.~\ref{fig:fit-results} shows the result of the simultaneous fit to the $m_{\mathrm{dijet}}$ distributions of the Signal and Control 
Regions, as well as the corresponding background-subtracted data distributions. 
The rather complex shape of the background invariant mass distribution results from the use of the six jet-based triggers, all of which have different 
jet \pt~thresholds and hence shape differently the invariant mass distributions. The fitted function models the data well, with a signal peak compatible with \zbb\ decays. 
The fitted signal yield is \mbox{$6420 \pm 640\,({\rm stat.})$} events. 

\section{Systematic uncertainties}
\label{sec:syst}
The sources of systematic uncertainties considered in this analysis, which may affect the fitted signal yield, the efficiency 
correction factor or both, are listed in Table~\ref{tab:syssummary}.
\begin{table*}[ht]
\begin{center}
\caption{The relative systematic uncertainties on the fitted yield of \zbb ,\NZ; the efficiency correction factor, \acc; and the fiducial cross-section, \zbbxsec, from each of the sources of uncertainty considered.} \vspace*{1mm}
\begin{tabular}{lccc}
\hline\hline
  Source of uncertainty & $\Delta$\NZ (\%) & $\Delta$\acc (\%) & $\Delta$\zbbxsec (\%)\\
\hline
& & & \\
Jet energy scale &  $+3.0/-1.5$ & $\pm$8.4 & $+6.5/-5.0$ \\
Jet energy resolution & $\pm5.3$ & $\pm0.2$ & $\pm5.1$\\
$b$-tagging & $\pm0.1$ & $\pm3.6$ & $\pm3.6$\\
Trigger modelling & N/A & $\pm 6$ & $\pm 6$ \\
Control Region bias & $+4.9/-5.5$ & N/A & $+4.9/-5.5$\\
Signal \split\ modelling & $\pm$0.9 & $\mp$2.0 & $\pm2.9$ \\
Signal $m_{\mathrm{dijet}}$ shape & $\pm$2.2 & N/A & $\pm$2.2\\
\zcc\ normalisation & $\pm$0.4 & N/A & $\pm$0.4 \\
\ttbar\ normalisation & $\pm1.2$ & N/A & $\pm1.1$\\
\wqq\ normalisation & $\pm1.0$ & N/A & $\pm1.0$\\
& & & \\
\hline\hline
\end{tabular}
\label{tab:syssummary}
\end{center}
\end{table*}

The jet energy scale (JES) and jet energy resolution (JER)
uncertainties are determined using the techniques described in Refs.~\cite{Aad:2011he, Aad:2012ag}. 
The JES uncertainty has a relatively large impact on the signal
efficiency, due to the \pt~requirements on the individual jets and the dijet system, but a
comparatively small impact on the 
fitted yield, because of the data-driven approach for the background determination and the fact that the location of the signal peak is
a free parameter of the EML fit. The JER uncertainty affects predominantly the fitted yield, since it modifies the MC-derived signal shape. 

The $b$-tagging efficiency in the simulation is scaled to reproduce the one in data and its uncertainty is evaluated by varying the data-to-MC scale factor applied
to each jet in the simulation within a range that reflects the
systematic uncertainty on the measured tagging efficiency for $b$-jets in ATLAS~\cite{ATLAS-CONF-2012-097, ATLAS-CONF-2014-004}. The \zcc~relative normalisation uncertainty is estimated in a similar way by varying the corresponding scale factors for charm jets in the simulation.

The uncertainty on \acc~due to a potential mis-modelling of the trigger efficiency is assessed using data events collected with a prescaled trigger that is fully efficient with respect 
to the analysis event selection. The full offline event selection is applied to these events and the efficiency for passing the analysis trigger requirements is compared to the
corresponding efficiency in the multi-jet MC sample, as a function of various kinematic variables. It is found that the two trigger efficiencies are consistent to within 6\%. Furthermore,
the trigger efficiency in the multi-jet MC sample, when considering only those events where the two $b$-tagged jets are labelled as true $b$-jets, is fully consistent with the trigger efficiency in the signal MC events. Based on these studies, a $\pm 6\%$ trigger efficiency modelling uncertainty is propagated to the cross-section measurement.

The uncertainty on the extracted signal yield due to potential differences in the background $m_{\mathrm{dijet}}$ shape between 
the Signal and Control Regions (``Control Region bias'') is assessed by repeating the EML fit for a range of \split~values
around the one used in the baseline selection to define the Control Region. These variations of the Control Region definition lead to small biases in the $m_{\mathrm{dijet}}$ shape relative to the Signal Region, resulting in non-zero slopes in the first-order polynomial fits to the distributions equivalent to the one in Fig.~\ref{fig:signal-vs-control}.
The non-zero slopes of these fits bracket the statistical uncertainty with which the slope of the first-order polynomial fit to Fig.~\ref{fig:signal-vs-control} is determined. The largest upwards and downwards variations in the fitted signal yield from the
EML fits following this procedure are propagated as systematic uncertainty to the cross-section measurement.

The impact on \acc\ of a possible mis-modelling of the distributions of the analysis selection variables, except \split, in the MC signal is
assessed by comparing the \sherpa~and {\sc{Pythia}} MC signal samples. It is found to be less than 1\% and therefore it is considered negligible. 
There is a 15\% discrepancy between the {\sc{Pythia}} and \sherpa~predictions for the efficiency of the Signal Region \split~requirement. Since the input variables to \split~depend primarily on the dynamics of $Z$-boson production, the
modelling by \sherpa~is tested by comparing a sample of events in the ATLAS 2012 data containing high-\pt~ 
$Z\rightarrow\mu^+\mu^-$ decays to a corresponding \sherpa~MC sample, with the dimuon system replacing the dijet system. 
The agreement is found to be very good, at the level of 2\%, and the residual
discrepancies are propagated as the ``Signal \split~modelling'' uncertainty on both \acc\ and the $R_Z$ fit parameter.
This uncertainty also covers the impact from possible differences between the PDFs used in the \sherpa\ signal sample and the data, given that the above $Z\rightarrow\mu^+\mu^-$ \sherpa\ sample uses the same PDF set as the \zbb\ \sherpa\ signal sample.

The difference obtained in the fitted signal yield when using the {\sc{Pythia}} signal model rather than the \sherpa\  one is used as an estimate of the
uncertainty on the measurement due to possible mis-modelling of the $m_{\mathrm{dijet}}$ shape in the MC signal. 

The impact on the measurement from the uncertainty on the \wqq~and \ttbar~normalisations is
assessed by varying the fixed number of events of each background in the Signal and Control Regions
independently by 50\% and repeating the EML fit.

\section{Results}
Using the extracted \zbb~yield, the estimated signal-efficiency correction factor and the integrated luminosity of the dataset,
the cross section in the fiducial region defined in Section~\ref{sec:xsec} is measured to be\vspace*{-5mm}

$$
  \sigma^{\mathrm{fid}}_{\PZ\rightarrow\Pbeauty\APbeauty} = 2.02\pm0.20\,({\rm
    stat.})\pm0.25\,({\rm syst.)}\pm 0.06\,({\rm lumi.}) \,\rm{pb}\,.
$$
The total systematic uncertainty is the result of adding in quadrature all the individual systematic uncertainties on
\zbbxsec~listed in Table~\ref{tab:syssummary}. 
It is further found that the signal $m_{\mathrm{dijet}}$ peak position is consistent with the \zbb\ expectation: 
$\delta M_Z=-1.5\pm 0.7\,(\mathrm{stat.})^{+3.4}_{-2.5}\,(\mathrm{syst.})\,\mathrm{\GeV}$. 
The good agreement with zero provides an independent confirmation of the good agreement between data and MC
on the energy scale of b-jets in ATLAS.

The robustness of the measurement is supported by several cross-checks and complementary studies. In particular, a consistent
cross-section measurement is obtained by applying a tighter $b$-tagging selection (with an efficiency of 60\% for tagging $b$-jets in 
a MC sample of \ttbar~events) or when the requirement on $\pt^{\mathrm{dijet}}$ is raised to 250\,\GeV\ or 300\,\GeV. Furthermore,
when the same methodology is repeated on two independent classes of events, those accepted by the dominant trigger described 
above and all other events, both measured cross sections ($1.99\pm 0.25\,(\mathrm{stat.})$\,pb and $1.87\pm 0.44\,(\mathrm{stat.})$\,pb
respectively) are fully consistent with the baseline measurement, even though 
the $m_{\mathrm{dijet}}$ distributions are significantly different in the two classes of events. 
In addition, when the background shape obtained from the baseline EML fit is used to fit for a signal in the sample of events with
 $0.45<$\,\split$\,<0.58$, the fitted signal yield in this sample is consistent with the number of signal events calculated based on the
measured cross section and the shape of \split~predicted by \sherpa. 
Finally, repeating the analysis with a number of alternative functional forms for the empirical description of the background 
shape (such as a log-normal function convolved with a fourth-order Bernstein polynomial) 
leads to negligible variations in the measured cross section compared to the systematic uncertainties of the measurement.

The measured cross section is compared to the particle-level, NLO-plus-parton-shower predictions of two MC generators, 
POWHEG and aMC@NLO, in the same fiducial region. In both cases, the cross section of the \mbox{$Z$ + 1-jet} process 
is calculated to NLO accuracy. 
For aMC@NLO, the $Z$ decay is simulated with MadSpin~\cite{Alwall:2011uj}. POWHEG is interfaced 
to {\sc{Pythia}} for parton showering, hadronisation and underlying-event
contributions, whilst aMC@NLO is interfaced to \herwigplus. The
particle-level predictions are then derived by applying to the generated events the fiducial selection defined in Section~\ref{sec:xsec}.  
The predicted cross sections are:\vspace*{-5mm}

$$
\mathrm{ POWHEG}:~~~ \sigma^{\mathrm{fid}}_{Z\rightarrow\Pbeauty\APbeauty} = 2.02~^{+ 0.25}_{-0.19} (\mathrm{scales})~^{+0.03}_{-0.04} (\mathrm{PDF})\, \mathrm{pb}~~
$$
$$
\mathrm{aMC@NLO}:~~~\sigma^{\mathrm{fid}}_{Z\rightarrow\Pbeauty\APbeauty} = 1.98~^{+ 0.16}_{-0.08} (\mathrm{scales}) \pm 0.03 (\mathrm{PDF})\, \mathrm{pb}\,.
$$
Both generators use the CT10 PDF set for the central value of the
prediction, and the renormalisation and factorisation scales are set to the \pt~of the $Z$ 
boson. The uncertainty due to the ambiguity in the renormalisation and factorisation scales is estimated by
doubling or halving them simultaneously. The PDF uncertainty is evaluated by varying the 52 PDFs in the CT10 NLO error set,
following the Hessian method and rescaling to the 68\% confidence level. 
Within the experimental and theoretical uncertainties, both predictions are completely consistent with the measured cross section.

POWHEG and aMC@NLO can also be used to provide an estimate of the fraction of the total cross section for \zbb~production at the LHC, 
with $\pt>200$\,\GeV, that is contained within the measured fiducial region. The ratio of the above cross sections to the cross sections calculated
without applying any particle-level requirements, only requiring $\pt>200$\,\GeV\ for the $Z$-boson before parton showering, is 0.53 for POWHEG and 
0.47 for aMC@NLO.

In conclusion, the high-\pt~\zbb~signal has been observed and its production cross section measured in a fully hadronic final state, in 
19.5\,fb$^{-1}$ of proton--proton collisions at $\sqrt{s}=8$\,\TeV\ recorded in 2012 by the ATLAS detector at the LHC.
Within the fiducial region defined in Section~\ref{sec:xsec}, the production cross section is measured to be\vspace*{-5mm}
 
$$
\sigma^{\mathrm{fid}}_{\PZ\rightarrow\Pbeauty\APbeauty} =2.02\pm 0.20\,(\mathrm{stat.})\pm0.25\,(\mathrm{syst.}) \pm 0.06\,(\mathrm{lumi.})\,\rm pb
$$
and is found to be in good agreement with the NLO-plus-parton-shower predictions from POWHEG and aMC@NLO. 

\section*{Acknowledgements}
We thank CERN for the very successful operation of the LHC, as well as the
support staff from our institutions without whom ATLAS could not be
operated efficiently.

We acknowledge the support of ANPCyT, Argentina; YerPhI, Armenia; ARC,
Australia; BMWF and FWF, Austria; ANAS, Azerbaijan; SSTC, Belarus; CNPq and FAPESP,
Brazil; NSERC, NRC and CFI, Canada; CERN; CONICYT, Chile; CAS, MOST and NSFC,
China; COLCIENCIAS, Colombia; MSMT CR, MPO CR and VSC CR, Czech Republic;
DNRF, DNSRC and Lundbeck Foundation, Denmark; EPLANET, ERC and NSRF, European Union;
IN2P3-CNRS, CEA-DSM/IRFU, France; GNSF, Georgia; BMBF, DFG, HGF, MPG and AvH
Foundation, Germany; GSRT and NSRF, Greece; ISF, MINERVA, GIF, I-CORE and Benoziyo Center,
Israel; INFN, Italy; MEXT and JSPS, Japan; CNRST, Morocco; FOM and NWO,
Netherlands; BRF and RCN, Norway; MNiSW and NCN, Poland; GRICES and FCT, Portugal; MNE/IFA, Romania; MES of Russia and ROSATOM, Russian Federation; JINR; MSTD,
Serbia; MSSR, Slovakia; ARRS and MIZ\v{S}, Slovenia; DST/NRF, South Africa;
MINECO, Spain; SRC and Wallenberg Foundation, Sweden; SER, SNSF and Cantons of
Bern and Geneva, Switzerland; NSC, Taiwan; TAEK, Turkey; STFC, the Royal
Society and Leverhulme Trust, United Kingdom; DOE and NSF, United States of
America.

The crucial computing support from all WLCG partners is acknowledged
gratefully, in particular from CERN and the ATLAS Tier-1 facilities at
TRIUMF (Canada), NDGF (Denmark, Norway, Sweden), CC-IN2P3 (France),
KIT/GridKA (Germany), INFN-CNAF (Italy), NL-T1 (Netherlands), PIC (Spain),
ASGC (Taiwan), RAL (UK) and BNL (USA) and in the Tier-2 facilities
worldwide.

\bibliographystyle{atlasBibStyleWithTitle}
\bibliography{ztobb}
\onecolumn
\begin{flushleft}
{\Large The ATLAS Collaboration}

\bigskip

G.~Aad$^{\rm 84}$,
B.~Abbott$^{\rm 112}$,
J.~Abdallah$^{\rm 152}$,
S.~Abdel~Khalek$^{\rm 116}$,
O.~Abdinov$^{\rm 11}$,
R.~Aben$^{\rm 106}$,
B.~Abi$^{\rm 113}$,
M.~Abolins$^{\rm 89}$,
O.S.~AbouZeid$^{\rm 159}$,
H.~Abramowicz$^{\rm 154}$
% ...
\bigskip
\\
% ...
$^{11}$ Institute of Physics, Azerbaijan Academy of Sciences, Baku, Azerbaijan\\
% ...
$^{84}$ CPPM, Aix-Marseille Universit{\'e} and CNRS/IN2P3, Marseille, France\\
% ...
$^{89}$ Department of Physics and Astronomy, Michigan State University, East Lansing MI, United States of America\\
% ...
$^{106}$ Nikhef National Institute for Subatomic Physics and University of Amsterdam, Amsterdam, Netherlands\\
% ...
$^{112}$ Homer L. Dodge Department of Physics and Astronomy, University of Oklahoma, Norman OK, United States of America\\
$^{113}$ Department of Physics, Oklahoma State University, Stillwater OK, United States of America\\
% ...
$^{116}$ LAL, Universit{\'e} Paris-Sud and CNRS/IN2P3, Orsay, France\\
% ...
$^{152}$ Institute of Physics, Academia Sinica, Taipei, Taiwan\\
% ...
$^{154}$ Raymond and Beverly Sackler School of Physics and Astronomy, Tel Aviv University, Tel Aviv, Israel\\
% ...
$^{159}$ Department of Physics, University of Toronto, Toronto ON, Canada\\
% ...
\end{flushleft}

\end{document}